\documentclass[printer]{aa}  

\usepackage{txfonts}

\usepackage{hyperref}
\usepackage{url}

\usepackage{graphicx}
\usepackage{xcolor}
\definecolor{forestgreen}{rgb}{0.10, 0.50, 0.10}
\renewcommand{\textbf}[1]{{\bfseries\textcolor{forestgreen}{#1}}}

\begin{document}

   
   \title{Interferometric Imaging  with LOFAR Remote Baselines \\
   of the Fine Structures of a Solar Type IIIb Radio Burst}
 \titlerunning{Interferometric imaging of Solar Type IIIb Burst with LOFAR}
 \authorrunning{Zhang et al}
   \author{PeiJin Zhang 
          \inst{1,2,3,4},
          Pietro Zucca
          \inst{4},
          Sarrvesh Seethapuram Sridhar
          \inst{4},
          ChuanBing Wang 
          \inst{1,2,3},
          Diana E. Morosan \inst{5},
          {Bartosz Dabrowski} \inst{6},
          {Andrzej Krankowski} \inst{6},
          {Mario M. Bisi} \inst{7},
          {Jasmina Magdalenic} \inst{8},
          Christian Vocks \inst{9},
          Gottfried Mann \inst{9}
          \thanks{Corresponding Author : Pietro Zucca (\email{zucca@astron.nl}), ChuanBing Wang (\email{cbwang@ustc.edu.cn})}
          }

   \institute{CAS Key Laboratory of Geospace Environment,
		School of Earth and Space Sciences, \\
		University of Science and Technology of China,
		Hefei, Anhui 230026, China
         \and
            CAS  Center for the Excellence in Comparative Planetology, USTC, 
    Hefei, Anhui 230026, China
        \and
        Anhui Mengcheng Geophysics National  Observation and Research Station, USTC, Mengcheng, Anhui 233500, China
        \and
        ASTRON, The Netherlands Institute for Radio Astronomy,
		Oude Hoogeveensedijk 4, 7991 PD Dwingeloo, The Netherlands
        \and
        {Department of Physics, University of Helsinki, P.O. Box 64, FI-00014 Helsinki, Finland}
        \and
        {Space Radio-Diagnostics Research Centre, University of Warmia and Mazury in Olsztyn, Olsztyn, Poland}
        \and
        {RAL Space, United Kingdom Research and Innovation (UKRI) - Science and Technology Facilities Council (STFC) - Rutherford Appleton Laboratory (RAL), Harwell Campus, Oxfordshire, OX11 0QX, U.K.}
        \and
        {Solar-Terrestrial Centre of Excellence—SIDC, Royal Observatory of Belgium, 1180 Brussels, Belgium}
		\and
		{Leibniz-Institut f\"ur Astrophysik Potsdam, An der Sternwarte 16, 14482 Potsdam, Germany}
             }

   \date{Received 2020-Apr-14; accepted ***}

 
  \abstract
   {Solar radio bursts originate mainly from high energy electrons accelerated in solar eruptions like solar flares, jets, and coronal mass ejections. A sub category of solar radio bursts with short time duration may be used as a proxy to understand the wave generation and propagation within the corona.}
   {Complete case studies of the source size, position and kinematics of short term bursts are very rare due to instrumental limitations. A comprehensive multi-frequency spectroscopic and imaging study was carried out of a clear example of a solar type IIIb-III pair.}
   {In this work, the source of the radio burst was imaged with the interferometric mode, using the remote baselines of the LOw Frequency ARray (LOFAR). A detailed analysis of the fine structures in the spectrum and of the radio source motion with imaging was conducted.}
   {The study shows how the fundamental and harmonic components have a significantly different source motion. The apparent source of the fundamental emission at 26.56\,MHz displaces away from the solar disk center at about 4 times the speed of light, while the apparent source of the harmonic emission at the same frequency shows a speed of $<0.02$\,c. The source size of the harmonic emission, observed in this case, is smaller than that in previous studies, indicating the importance of the use of the remote baselines.}
   {}


   \maketitle
%

\section{Introduction}
\label{sec:intro}
	
	Solar type III radio bursts are among the brightest solar radio bursts observed \citep[see, e.g.][]{wild1963solar, Reid2014RAA}. The major defining characteristic of type III radio bursts is their fast frequency drift rate, which is about 10\,MHz/s in the decametre range \citep{melnik2011observations, zhang2018type}. Type III bursts are generated by weakly relativistic electron beams moving through the coronal plasma along open magnetic field structures. It is generally believed that the electron beam can excite Langmuir waves at the local plasma frequency, which can be expressed as $f_{pe}[\rm{kHz}] \approx 8.98 \sqrt{N_e[\rm{cm^{-3}}]}$, the Langmuir wave can than be converted into electromagnetic waves at the fundamental or harmonic frequency through non-linear processes \citep{ginzburg1958possible,Reid2014RAA}. One alternative mechanism, namely the electron cyclotron maser (ECM) emission was also proposed for the excitation of type III bursts \citep{Wu2002, Chen2017}. 
	
	A number of type III bursts show fundamental\,--\,harmonic (F\,--\,H) pair structure. Type IIIb-III pair is a special kind of type III F\,-\,H pair events, which are characterised by the presence of fine structures in the fundamental part \citep{DulkTypeIIIPol1980A&A, Melnik2018}. The generation mechanism of the type IIIb resulting in these fine structures in the spectrum is still debated. Base on the ECM emission, the fine structures can be generated due to modulation of the wave excitation by low-frequency MHD waves \citep{wang2015scenario,zhao2013ApJ}. The commonly held belief about the plasma emission mechanism is that density inhomogeneities in the background plasma create a clumpy distribution of Langmuir waves and are the cause of type IIIb fine structures \citep{Takakura1975,Kontar2001,Loi2014}.  \cite{Mugundhan2017IIIbdensity} used the flux intensity variation at different frequencies to estimate the amplitude and scale of the density fluctuations. Most of the previous type IIIb bursts studies focused on the dynamic spectrum. 
	
	Imaging of the type IIIb source can help the understanding the generation and propagation of the radio wave. \cite{Abranin1976Angular} used Ukrainian T-shaped Radio telescope (UTR-2) operated as one-dimensional heliograph to study the angular size of the type IIIb-III pair events. They studied the source size at 25\,MHz and 12.5\,MHz, and found that there are no significant changes in the second harmonic part, the source size for the harmonic part is about 20 arcmin and $<$40 arcmin at 25 and 12.5\,MHz, respectively. The source size can be smaller than 10 arcmin for the fundamental part at 25\,MHz. \cite{kontar2017imaging} performed tied array imaging and spectroscopy with core baselines {(3.5 km)} of a type IIIb event with the beamformed mode of the LOw Frequency ARray (LOFAR) \citep{van2013lofar}. The study showed that, at a given frequency along the burst striae, the speed of the apparent source of the fundamental part moves faster than the harmonic part, and the source size (or area) increasing rate of the fundamental part is significantly larger than the harmonic part. The study also indicated that the observed wave duration and positions are convoluted by the wave generation, the scattering and refraction of the wave during the propagation. \cite{Sharykin2018lofar} used the LOFAR beamformed method and studied the source size and position along several striae. The result shows that the source position along each stria is moving away from the Sun center, and the area of the source is increasing with time. 
	
	The short-term narrow-bandwidth radio emission is believed to be generated from a small area, while the observed source can have a large size and moving speed. For a better comprehension of the factors that influence the visual source, we acquire high spatial resolution images for the fine structures in the radio emission.
	
    In this work, we image for the first time a type IIIb-III pair event with LOFAR interferometric observation of the core and remote stations. This paper is organized as follows: In Section 2,  {the observation mode, data reduction process and dyanmic spectrum of this type IIIb-III pair are described briefly}. In Section 3, {different source properties of the fundamental and harmonic components are presented. Finally, we discuss our results and summarize our conclusions in Section 4}.

\section{Observations and data reduction}

{LOFAR is an advanced radio antenna array which operates in the 10\,--\,250~MHz frequency range. LOFAR consists of two type of antennas - the Low Band Array (LBA) and the High Band Array (HBA) - which are sensitive to the 10\,--\,90~MHz and the 110\,--\,250~MHz frequency ranges, respectively. LOFAR has 52 stations, 38 of which are located in the Netherlands, and 14 international stations in Germany, Poland, France, Sweden, Ireland, the UK, and Latvia. Of the 38 Dutch stations, 24 stations are densely packed and are referred to as the ``core'' stations. The remaining 14 ``remote'' stations are sparsely distributed across northern Netherlands. The core and the remote stations provide excellent instantaneous \textit{uv}-coverage to produce images at high spatial resolution combined with good sensitivity to emissions on large angular scales. This makes LOFAR an excellent instrument to study a complex source like the Sun. In this work, we make use of data from only the Dutch stations. Owing to its flexible software backend, it can observe in different modes including standard interferometric imaging, tied-array beam-forming, and real-time triggering on incoming station data-streams.}

{We carried out a simultaneous interferometric and beamformed observations of the Sun with the Dutch LOFAR array using its LBA on 13 April 2019 (project code: LT10\_002). The core and the remote  stations were used in the interferometric mode while only the core stations were used to obtained the beam-formed mode. In the interferometric mode, we used 60 non-contiguous subbands (SBs) to cover the 10\,--\,90~MHz frequency range. Each SB has a bandwidth of 195.3~kHz and is further sub-divided into 16 channels resulting in a frequency resolution of 12.2~kHz. We recorded the visibility data with a correlator integration time of 0.167~s. Using the same observational setup, we also simultaneously observed Taurus~A as a calibrator to derive the station gains. The projected baseline length for our observation is about 80\,km which results in a theoretical spatial resolution of about 36 arcsec. However, since we make images of the Sun with high temporal and frequency resolution, the achieved spatial resolution is higher due to the instantaneous \textit{uv}-coverage and the applied visibility weighting.}

{We calibrated our correlated visibility data using the Default Pre-Processing Pipeline \citep[DPPP;][]{van2018dppp}. We derived the amplitude and the phase solutions using a model\footnote{\url{https://github.com/lofar-astron/prefactor/blob/master/skymodels/Ateam_LBA_CC.skymodel}} for Taurus A and applied the derived gain solutions to the solar dataset. While deriving the gain solutions, we accounted for the difference in the station beam response towards the Sun and Taurus A. We imaged the calibrated visibilities using WSClean \citep{offringa2014wsclean, offringa2017} making use of it multiscale deconvolution algorithm. While imaging, we weighted the visibilities using the Briggs weighting scheme with the robust parameter set to 0.2 \citep{briggs1995}. We imaged and deconvolved visibility data from each 0.168\,s separately, which allowed us to resolve the temporal variations in the Sun.}

	\begin{figure}
		\centering
		\includegraphics[width=0.95\linewidth]{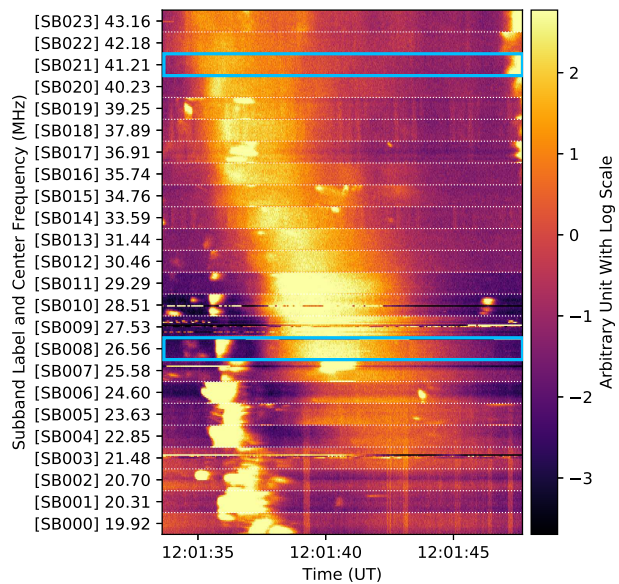}
		\caption{The dynamic spectrum of the type IIIb-III pair event, it contains 24 Sub-bands, the band-width of each Subband is about 0.183\,MHz. The frequency of this "dynamic spectrum" is not continuous. The blue box highlighted the Subband SB008 and SB021, which is used in the interferometric imaging.}
		\label{fig:0}
	\end{figure}

	\begin{figure*}[h] 
	\centering
	\includegraphics[width=0.75\linewidth]{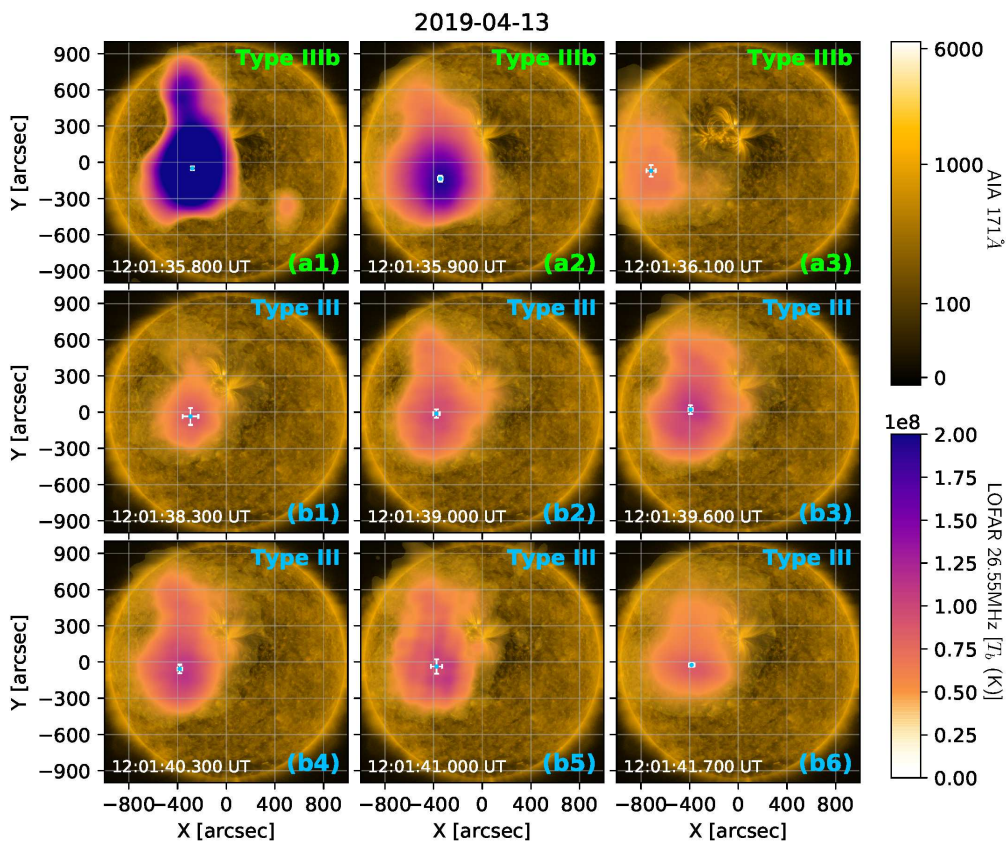}
	\caption{The interferometry observation of the type IIIb-III pair. Panel (a1\,-\,3) is the interferometric image of the  fundamental (type  IIIb) part, and  Panel (b1\,-\,6) shows the second harmonic (type III) part in subband SB008 (26.56\,MHz). The radio flux intensity is overlaid on the EUV image observed by SDO/AIA at 12:01:33\,UT. The unit of radio flux intensity in this is converted to brightness temperature (Kelvin). The times of snapshots is marked {by vertical blue lines} in the left-top sub-panel of Figure \ref{fig:2}. The peak position of the radio flux intensity is marked as blue dots with position uncertainty as white errorbar in each snapshot.}
	\label{fig:1}
	\end{figure*}

{From our observations of the Sun, we identified the type IIIb-III event on 13 April 2019 at 12:01 UT (see Fig.~\ref{fig:0}). From Figure \ref{fig:0} we can see that this is a typical type IIIb-III pair event. There is striae structure in the fundamental part.} The frequency ratio of the harmonic and fundamental band is about 1.6.  The upper frequency limit of the fundamental (type IIIb burst) part is about 30\,MHz. We selected two subbands, with central frequencies 26.56 and 41.21\,MHz, which are relatively free of other bursts. These subbands are indicated with blue boxed in Fig~\ref{fig:0}.


	\section{Results}
	
	Figure \ref{fig:1} shows the LOFAR interferometric imaging of the Sun at 26.56\,MHz overlaid  on the EUV image observed by SDO/AIA at 12:01:33\,UT \citep{SDOAIA}. We see from Figure \ref{fig:1} (a1-3) that, for the fundamental part, the apparent source moved {about 500 arcsec} within 0.3 seconds {toward the limb}. While, for the harmonic part shown in Figure \ref{fig:1} (b1-6), the  apparent source position is stable for the 3.4 seconds as shown in these six frames, {the attached movie shows the variation of source size and position of fundamental and harmonic emission}. At 26.56\,MHz, the source of the type IIIb has a higher brightness temperature, and moves faster than the source of the type III. Figure \ref{fig:1} also shows that, by employing long Dutch LOFAR baseline, we can resolve the source shape with unprecedented spatial resolution. The source shape in Figure \ref{fig:1} (a1,2) and (b2-4) are similar. 
	
\begin{figure*}[h]
		\centering
		\includegraphics[width=0.42\linewidth]{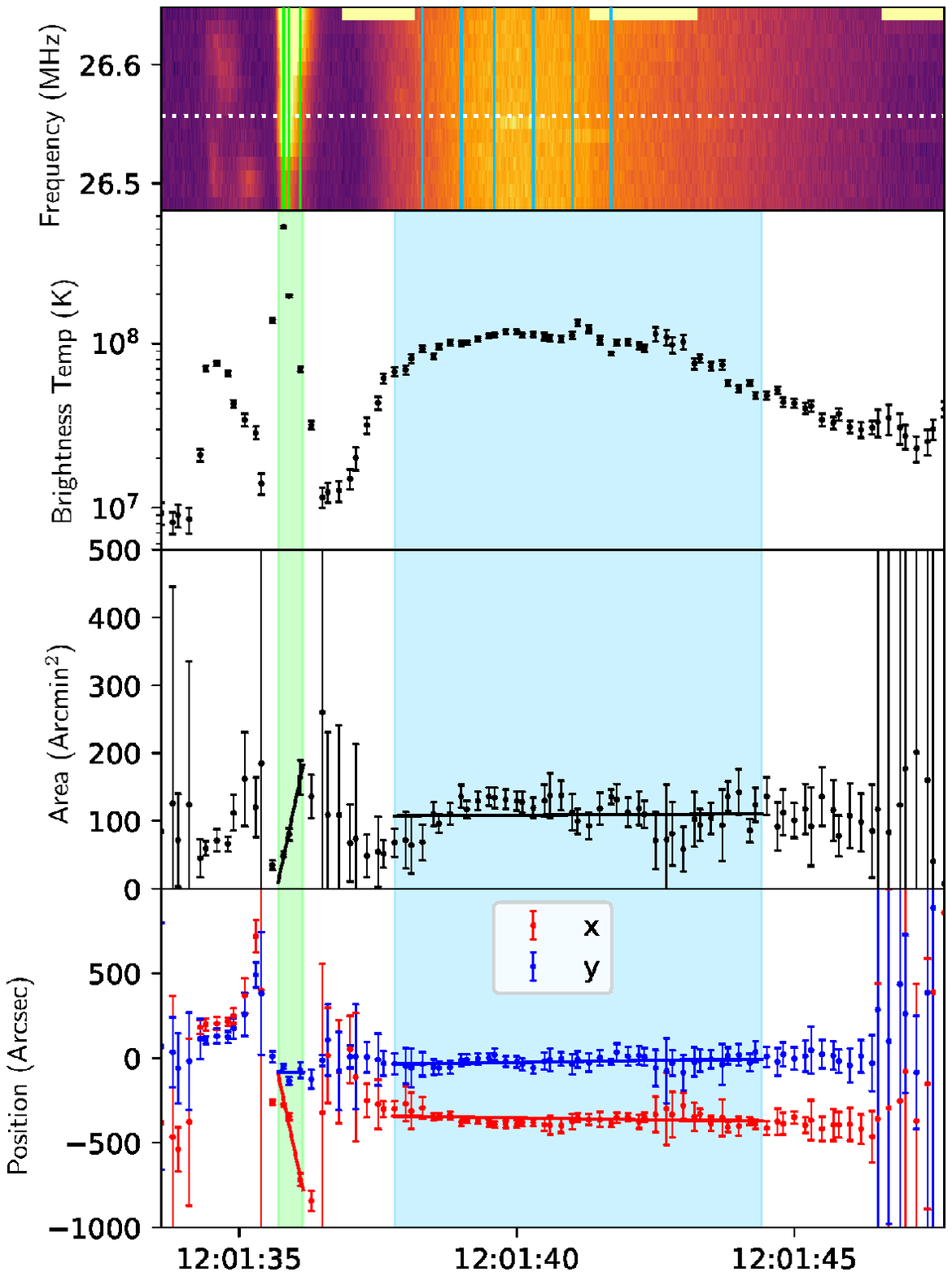}
		\includegraphics[width=0.42\linewidth]{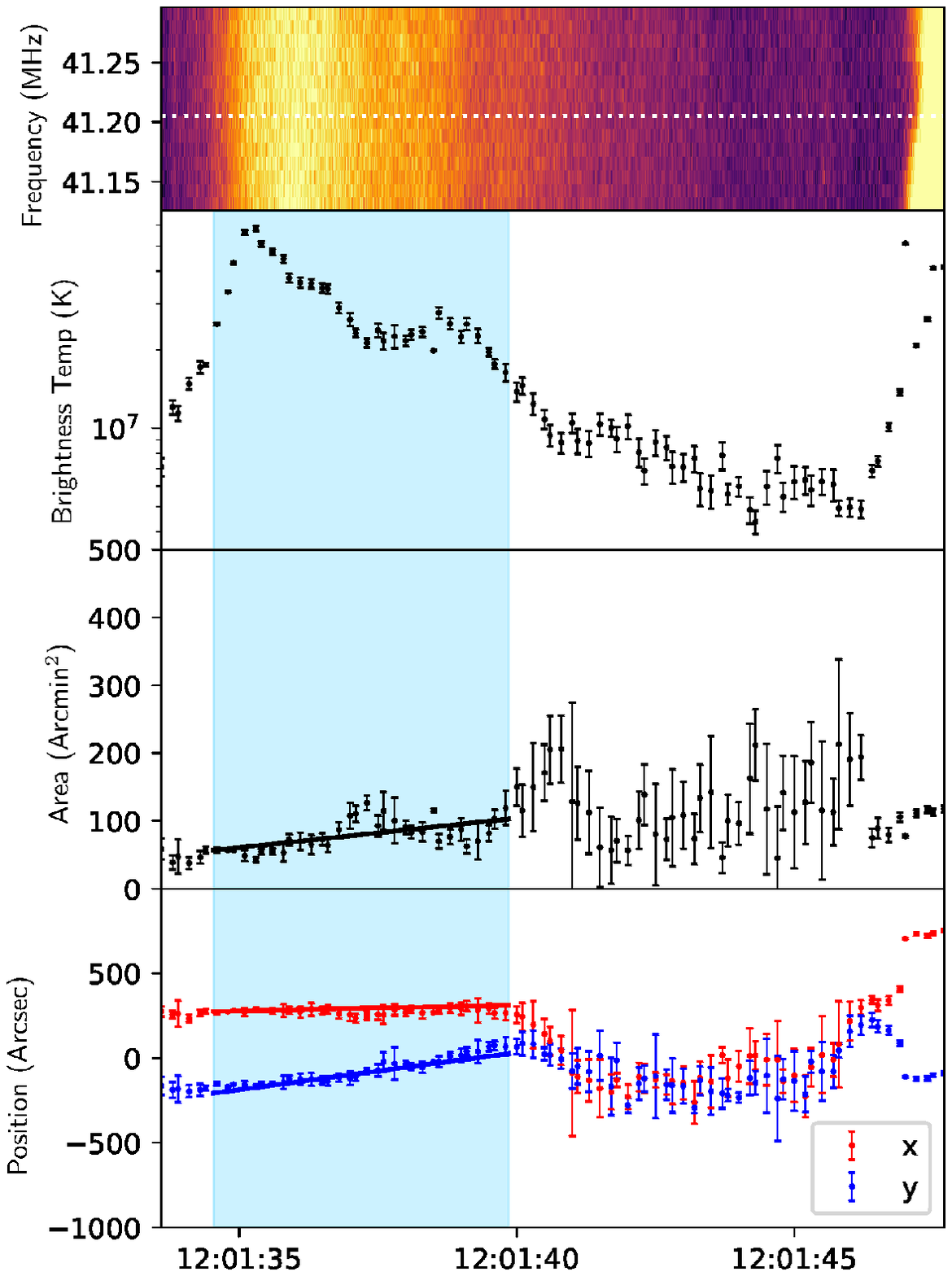}
		\caption{The dynamic spectra of Subband SB008 (left) and SB021 (right) with the center frequency of 26.56\,MHz and 41.21\,MHz (top panel), and the source brightness temperature (2nd panel), the size and position of these two subbands (3rd and 4th panel). The fundamental part is marked as green shadow, the harmonic part is marked as blue shadow, and the errorbar marks the uncertainty.}
		\label{fig:2}
	\end{figure*}

	{The brightness temperature was obtained from the flux intensity with the unit of [Jy/Beam] with $ T_b = {\lambda^2 S}/{(2k\Omega)}  $, where $\lambda$ is the wavelength, $k$ is the Boltzmann constant, $S$ is the flux intensity, $\Omega$ is the solid angle of the beam. The uncertainty of the brightness temperature is determined by the standard deviation of the residual map. The source position and it's uncertainty is obtained with the  two-dimensional Gaussian fit method \citep{kontar2017imaging,zhang2019source}. The  source area ($A_s$) is determined by the FWHM of the brightness temperature map. The uncertainty of the source is determined as $\delta A_s = 2 \sqrt{A_s (\delta x^2 + \delta y^2 )} $, where $\delta x$ and $\delta y$ is the uncertainty of the source position.} We did a survey on the source brightness, size, and movement for two Subbands, namely 26.56\,MHz and 41.21\,MHz, shown in Figure \ref{fig:2}. The fundamental and harmonic parts both appear in 26.56\,MHz (shown in the left panel of Figure \ref{fig:2}), the Subband of 41.21M\,Hz only contains the harmonic part (shown in the right part of Figure \ref{fig:2}). The statistical and linear-fit results are shown in Table \ref{tab:1}.

	\begin{table*}[h]
		\centering
		\caption{The FWHM duration, initial area, area increasing rate, and apparent velocity of the radio source in the two selected subbands, the velocity is in the unit of the speed of light. The second column is the results of the fundamental part of Subband 26.56\,MHz, the third and fourth column are the results of harmonic part of Subband 26.56\,MHz and 41.21\,MHz, respectively.  \label{tab:1}}
		\begin{tabular}{cccc}
			\hline
			                            & 26.56\,MHz (F)       & 26.56\,MHz (H) & 41.21\,MHz (H)\\
			\hline\hline
			$t_{FWHM}$ [s]              & 0.45              & 6.61        & 5.31        \\
			$A_0$   [arcmin$^2$]        & 50.7 $\pm$ 3.7    & 67.7$\pm$ 20.6        & 56.7 $\pm$ 3.8        \\
			$dA/dt$ [arcmin$^2$/s]      & 382 $\pm$ 30      & 0.4 $\pm$ 2.0         & 8.7 $\pm$ 2.1         \\
			$v_x$    [c]                & -3.67 $\pm$ 0.72  & -0.010 $\pm$ 0.007     & 0.002 $\pm$ 0.004        \\
			$v_y$   [c]                 & -0.03 $\pm$ 0.71  & 0.010 $\pm$ 0.006      & 0.115 $\pm$ 0.006   \\
			\hline    
		\end{tabular}
	\end{table*}
	 
	{Figure \ref{fig:2} and Table \ref{tab:1} shows the inteferometric analysis} of the type IIIb-III event in 26.56 and 41.21\,MHz with LOFAR core and remote stations. The flux peak time of the harmonic {emission} in 41.21\,MHz is close to the peak time of the fundamental {emission} in 26.56\,MHz. The main results of the imaging is as follows: 
	
	\begin{itemize}
	    \item For the frequency of 26.56\,MHz, the source area of the fundamental {emission} increase from about 50\,$\rm{arcmin^2}$ to 200\,$\rm{arcmin^2}$ within 0.45 seconds, while the source area of the fundamental is stable near about  100\,$\rm{arcmin^2}$ for the 6.61 seconds of duration.
	   
	    \item For the frequency of 41.21\,MHz, only harmonic {emission} is observed. The visual source area increase from about 50\,$\rm{arcmin^2}$ to 100\,$\rm{arcmin^2}$ in 5.31 seconds.
	    
	    \item The visual speed of the source of fundamental {emission} at 26.56\,MHz is about four times the speed of light. 
	    
	    \item The visual speed of the sources of harmonic {emission} at 26.56\,MHz and 41.21\,MHz are less than 0.11 times the speed of light.
	\end{itemize}

	There are significant differences between the source properties of the fundamental and harmonic waves. For the frequency subband of 26.56\,MHz, the source area of the fundamental part increases fast, the increasing rate is 382 arcmin$^2$ per second starting from about 50 arcmin$^2$. While, the the source area increasing rate of the harmonic part is nearly zero for 26.56\,MHz, and 8.7\,arcmin$^2$ per second for 41.21\,MHz. The speed of the apparent source is also widely divergent between the fundamental and harmonic part. We need to note that, the speed of the apparent source is the visual speed of the brightest point in single frequency, not the physical position of the electron beams. The apparent source speed of the fundamental part is about $3.7$\,c for emissions at 26.56\,MHz. While source position of the harmonic part is stable, and the speed is about 0.015\,c for 26.56\,MHz and 0.11\,c for 41.21\,MHz.

	\section{Discussion}
	
	The source size observed in this case is smaller than the results of previous study.  The FWHM area of the source is about 100\,$\rm{arcmin^2}$ for 26.56 and 41.21\,MHz, {with a minimal value of 50\,$\rm{arcmin^2}$}.  \cite{Abranin1976Angular} measured the one dimensional angular width of the type III radio burst source with UTR, they obtained about 300\,$\rm{arcmin^2}$ for the source of 26\,MHz. \cite{kontar2017imaging} used beam-formed imaging of LOFAR core station and analyzed the variation of the source size and position of the type IIIb-III pair, the source area is about 500\,$\rm{arcmin^2}$ for the fundamental and 700\,$\rm{arcmin^2}$ for the harmonic at 32.5\,MHz. \cite{Sharykin2018SoPhyLOFAR} obtained about 350\,$\rm{arcmin^2}$ for the straie near 30.1\,MHz with the LOFAR beam-formed observation of type IIIb. The linearly-fitted expansion rate of the fundamental wave in this case is 382\,$\rm{arcmin^2/s}$ , which is larger than previous results. \cite{kontar2017imaging} obtained the 180\,$\rm{arcmin^2/s}$ for fundamental, and 50\,$\rm{arcmin^2/s}$ for harmonic {emission}. \cite{Sharykin2018SoPhyLOFAR} measured the expansion rate along the straie of the type IIIb burst, the value of the expansion rate varies between 50 and 200\,$\rm{arcmin^2/s}$ during the burst. 
	
	{The observed size and movement of the source is determined by both the original source property at the wave excitation site, namely the "real source" and the propagation effect of the radio waves.} The type III radio burst is excited by the electron beam propagating outward along open magnetic field lines. The {size and position of the real source} at a given frequency $f_s$ {is determined by the wave excitation condition and the cross section of the electron beam at the altitude where the local plasma frequency satisfies $f_{pe} = f_{s}$ (or $f_s/2$ for harmonic) according to the plasma emission mechanism \citep{ginzburg1958,Reid2014RAA}}. {The background conditions, namely the electron density and} the magnetic fields are stable within the time scale of second. Consequently, the size and position of the real source  {should be} stable at the generation site for the wave of $f_s$ {for short term bursts}. While, the observation in this case shows that {the observed source of the fundamental wave changed significantly}. This indicates that {the observed radio source is largely influenced by the  wave propagation effects, namely the wave refraction and the scattering}. {For these two factors, the variation of the source size is mainly dominated by the scattering effect. According to the simulation of anisotropic scattering \citep{kontar2019arXiv190900340K}, the source FWHM width due to the scattering could be 1 Solar radius at 32\,MHz for fundamental waves with anisotropy parameter $\alpha$ = 0.3 and a level of turbulence $\epsilon$ = 0.8.}  
	
\begin{figure}[h]
		\centering
		\includegraphics[width=0.7\linewidth]{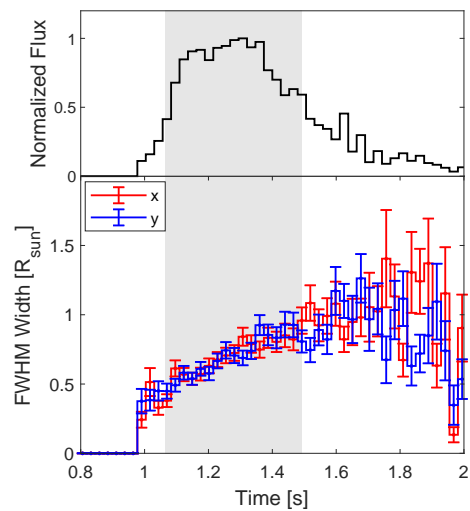}
		\caption{The simulation result of the observed flux (upper panel) and the apparent source size (width) along the $x$ and $y$ axis (lower panel) for a impulse point source {{with frequency 26~MHz located at the solar disc center}}. In the simulation, the turbulence level of $\epsilon=0.3$ and an-isotropic parameter $\alpha=0.3$. The gray area represents the time range of the Flux FWHM.}
		\label{fig:3}
	\end{figure}

	{The observed source expansion at a given frequency corresponds to the variation of the amount of scattering experienced by the waves. In this case, the source size of the fundamental emission increase from 50\,$\rm{arcmin^2}$ to 200\,$\rm{arcmin^2}$, which implies that the radio waves observed at the beginning of the type IIIb burst experience less scattering than that at the end of the type IIIb burst. Using the anisotropic simulation model by \cite{kontar2019arXiv190900340K}, we simulated the source expansion process for an ideal pulse point source of the fundamental emission at $f_s=26$\,MHz. The results are shown in Figure \ref{fig:3}. One can see that the source size increases during the time range of flux FWHM. In addition, the minimal observed source size can set a lower limit to the scattering effect. With LOFAR remote baseline for the imaging, the average fitted beam-size is 180\,$\rm{arcsec}$ (3\,$\rm{arcmin}$), which is also the resolution limit of this method. In this event, the minimal observed source size of fundamental emission is 50\,$\rm{arcmin^2}$ for the fundamental wave at $f_s=26$\,MHz, which requires $\epsilon=0.3$ for the background according to the simulation result shown in Figure \ref{fig:3}. Of course, statistical work of more type IIIb-III pair events analyzed with the interferometric of LOFAR remote baseline are needed to constrain the parameters related to the wave scattering.}

	{Both the wave refraction and the scattering can contribute to the high speed visual movement of the source. In general, the refraction displaces the observed source position inward from its real location, and the scattering produces outward displacement of the apparent source.} Ray-tracing simulation results show that, for an ideal point pulse source, the scattering and refraction of the waves will result in a visual motion of the source toward the solar center with time  \citep{arzner1999radiowave}, {though the transient position of the source is still been displaced outward from its real position}. While in this case, the observed source is moving outward from the solar disk center. {This may be due to the simple assumptions used in the ray-tracing simulation, such as an ideal pulse point source,  sphere symmetric distribution of the background density, and isotropic scattering etc. In fact, the corona is a highly non-uniform medium with a number of discrete large structures, including loops, coronal holes, helmet streamers, bright and dark rays etc.}
	
	{Some observations suggest that type III burst may be generated in regions of enhanced density or along the streamers \citep{Fainberg_Stone1974}. As a result, the radio waves will propagate away from the density-enhanced region after excitation, in an arbitrary direction depending on the observation view angle with respect to the streamer.}  On the other hand,  \cite{duncan_wave_1979} and \cite{calvert_wave_1995} suggested that the wave of the radio burst can be guided by the under-dense flux tube. The escape point of the wave from the tube is determined by {the density depletion factor, the wave frequency,}  and the wave angle (the angle between the wave vector and the magnetic field).  {For the event in this work, the harmonic component has a relatively stable apparent position and expands slowly. This indicates the ducting effect is weak for harmonic waves, or in other words the tube depletion factor is low. For the fundamental part, we expect that waves with larger wave angle could escape earlier from the tube at a lower altitude than that of the waves with smaller wave angle from the same generation site. After the waves escaped from the flux tube, the scattering will displace the wave source outward further. Moreover, the cross section of the tube increases with height due to the divergence of magnetic field, also producing an increase of the wave source size. This may explain the  large expansion rate of the type IIIb source size, accompanying with the super-luminal source motion observed in this event. Ray-tracing simulations with photon scattering in different background environments may be help to clarify these in the future. }

\begin{acknowledgements}
The analysis of the SDO/AIA data is done with Sunpy, gnu Parallel is used for the acceleration of the imaging process. We are thankful to the ASTRON/JIVE Summer Student Programme 2019 for the financial support. This paper is based on data obtained with the International LOFAR Telescope (ILT) under project code LT10 002. LOFAR \cite{van2013lofar} is the Low Frequency Array designed and constructed by ASTRON. It has observing, data processing, and data storage facilities in several countries, that are owned by various parties (each with their own funding sources), and that are collectively operated by the ILT foundation under a joint scientific policy. The ILT resources have benefitted from the following recent major funding sources: CNRS-INSU, Observatoire de Paris and Universite Orleans, France; BMBF, MIWF- NRW, MPG, Germany; Science Foundation Ireland (SFI), Department of Business, Enterprise and Innovation (DBEI), Ireland; NWO, The Netherlands; The Science and Technology Facilities Council, UK; Ministry of Science and Higher Education, Poland.
The research in USTC was supported by the National Nature Science Foundation of China (41574167 and 41974199) and the B-type Strategic Priority Program of the Chinese Academy of Sciences (XDB41000000).
\end{acknowledgements}

%
%
\bibliographystyle{aa}
\bibliography{citebib}

\end{document}